\documentclass[aps,prd,twocolumn,10pt,superscriptaddress,amssymb,amsmath,nofootinbib]{revtex4-2}

\usepackage[utf8]{inputenc}
\usepackage{lmodern}
\usepackage[T1]{fontenc}
\usepackage{graphicx} % Required for inserting images
\usepackage[pdftex,breaklinks,colorlinks,
linkcolor=Blue,
citecolor=teal,
anchorcolor=red,
urlcolor=cyan,
pdfencoding=auto]{hyperref}
\usepackage[dvipsnames]{xcolor}
\usepackage{mathtools}
\usepackage{mathrsfs}
\usepackage[scr=boondoxo]{mathalpha}
\usepackage{booktabs}
\usepackage{multirow}
 \usepackage{array}
    \newcolumntype{M}{>{$}c<{$}} % Defines a new column type 'M' for centered math

\usepackage{bm}
\usepackage{float}

\usepackage{nicefrac}


\newcommand{\beeq}{\begin{equation}}
\newcommand{\eeq}{\end{equation}}

\definecolor{myYello}{HTML}{E8A317}

\newcommand{\orcid}[1]{\href{https://orcid.org/#1}{\includegraphics[width=8pt]{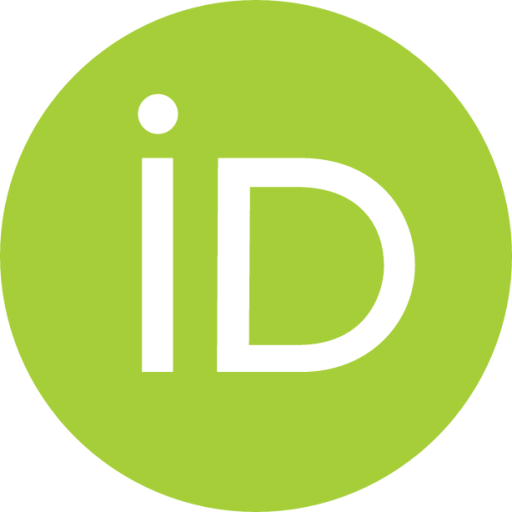}}}

\begin{document}

\title{Relativistic signatures of scalar dark matter in extreme-mass-ratio inspirals}

\author{Robrecht Keijzer \orcid{0009-0002-4580-3651}}
\email{robrecht.keijzer@kuleuven.be}
\affiliation{Institute for Theoretical Physics, KU Leuven, Celestijnenlaan 200D, 3001 Leuven, Belgium}
\affiliation{Leuven Gravity Institute, KU Leuven, Celestijnenlaan 200D, 3001 Leuven, Belgium}

\author{Simon Maenaut \orcid{0000-0003-1464-2605}}
\affiliation{Center of Gravity, Niels Bohr Institute, Blegdamsvej 17, 2100 Copenhagen, Denmark}

\author{Henri Inchauspé \orcid{0000-0002-4664-6451}}
\affiliation{Institute for Theoretical Physics, KU Leuven, Celestijnenlaan 200D, 3001 Leuven, Belgium}
\affiliation{Leuven Gravity Institute, KU Leuven, Celestijnenlaan 200D, 3001 Leuven, Belgium}

\author{Thomas Hertog \orcid{0000-0002-9021-5966}}
\affiliation{Institute for Theoretical Physics, KU Leuven, Celestijnenlaan 200D, 3001 Leuven, Belgium}
\affiliation{Leuven Gravity Institute, KU Leuven, Celestijnenlaan 200D, 3001 Leuven, Belgium}

\begin{abstract}
    We study gravitational wave emission by circular extreme-mass-ratio systems in a spherically symmetric scalar environment. Previous studies have focused on the impact of scalar radiation channels, revealing a rich structure of resonances, sharp features and floating orbits. Through the backreaction of the cloud on the metric, corrections to the gravitational sector come in at the same order. We develop the computational methods, and provide a characterization of this new, fully relativistic cloud signature. Remarkably, corrections to the polar sector can dominate all other dissipative corrections. We identify scalar field masses $M\mu\lesssim 0.12$ as the regime where polar can overtake axial and scalar channels at small separation. For small $M\mu$, vacuum dephasing is dominated mostly by conservative and polar cloud corrections, with scalar radiation acting as only a minor correction. At large $M\mu$, both terms terms are shown to be highly non-negligible. Our results therefore motivate including these relativistic signatures in beyond-vacuum EMRI templates.
    
\end{abstract}

\maketitle

%\tableofcontents
\textit{\textbf{Introduction.}} 
Over the last decade, gravitational wave (GW) astronomy has opened a precision window onto compact object dynamics in the strong gravity regime \hbox{\cite{Abbott_2016,Catalog2,Roadmap}}. In vacuum, General Relativity predicts that astrophysical, stationary black holes are uniquely characterised by their mass and spin \cite{Israel,Carter,Robinson}. Next-generation GW detectors bring the exciting possibility to challenge this simplified picture with unprecedented accuracy \cite{Roadmap,Roadmap2,LISA4}. Intermediate and extreme-mass-ratio inspirals (IMRIs/EMRIs) are especially promising probes \cite{LISA1,LISA2,LISA3,LISA4,EMRI}, uniquely accessible in the milliHertz frequency range of space-based detectors, such as the recently adopted LISA mission \cite{LISA4}, or Tianqin \cite{Tianqin,Tianqin2} and Taiji \cite{Taiji}. Their large mass asymmetry allows the secondary to probe the strong gravity regime in a narrow frequency band for years, making subtle cumulative effects measurable \cite{EMRI,FEW1,FEW2}. Environmental imprints from accretion disks \cite{accr1,accr2,accr3,accr4,accr5,accr6,accr7,accr8,accr9,accr10,accr11,accr12,ConorTorque}, a third perturber \cite{Perturb,Perturb2,Perturb3}, dark-matter structures \cite{accr6,accr9,Mitra_2025,POSTADIABA,DM1,DM2,DM3,DM4,DM5,DM6,DM7,DM8,DM9,DM10,DM11,DM12,DM13,DM14}, and ultralight scalar fields \cite{scal1,scal2,scal3,scal4,scal4.5,scal5,scal6,scal6.5,scal8,scal9,scal10,scal11,scal12,scal12.5,scal13,scal14,BritoMain,Duque_2024,scal17,Dyson,Khalvati_2025,scal15,scal9.5,scal16} have been studied. Ultralight scalar fields are an especially exciting prospect: they can assemble dense `scalar clouds' through superradiance around spinning black holes \cite{Detweiler,Superradiance,Validity_expansion,SR1}, are predicted by extensions to the standard model and cosmology \cite{StringTheory,FuzzyDM,FuzzyDM2,FuzzyDM3}, and allow a direct measurement of the dark matter field properties through only gravitational interactions \cite{scal10,BritoMain,Khalvati_2025}.

\smallskip

Recent relativistic analyses have mapped scalar radiation for EMRIs embedded in a scalar cloud \cite{BritoMain,Duque_2024,Dyson,scalarRadiation}, showing that it may rival GW losses in parts of parameter space. Sharp resonances and floating orbits could be a smoking gun signature of beyond vacuum GR deviations \cite{Floating_orbits,scal9,BritoMain}. However, as noted in Ref. \cite{BritoMain} but not included as a correction in Ref. \cite{Duque_2024,Dyson,scalarRadiation}, corrections through the backreaction of the cloud on the metric come in as a fully relativistic effect at the same order. For spherically symmetric clouds, leading corrections to \emph{axial} GW fluxes mostly decouple, and reduce to a gravitational redshift effect \cite{BritoMain, Cardoso_2022}. This metric coupling is found to give a subdominant, but possibly non-negligible correction to scalar radiation \cite{BritoMain}. In this work, we perform the first study on the fully coupled polar sector in a scalar cloud environment. Remarkably, corrections to the polar sector of the GW flux can \textit{dominate} all other dissipative cloud signatures in the waveform. As a function of the scalar field mass, we identify $M\mu \lesssim 0.12$ as the regime where the polar term overtakes axial and scalar channels at small separation, while remaining non-negligible at higher $M\mu$. The new signature is well described by a redshift for $M\mu \lesssim0.08$ at small separations. Comparing the dephasing caused by each correction identifies a significantly more involved picture than purely dissipative scalar radiation. Conservative corrections can further dominate flux corrections, both in the high and low $M\mu$ regime. Our results therefore motivate including polar and conservative corrections in beyond-vacuum EMRI templates, with the possibility to significantly alter existing estimates of cloud parameter estimations.

\smallskip
\textit{\textbf{Framework.}} Following recent developments on the multi-parameter formalism to consistently include environmental effects, we can model both the secondary and the environment perturbatively on the Schwarzschild black hole background \cite{BritoMain,Datta,Dyson,ConorTorque,scalarRadiation,Duque_2024}. At leading order in the mass ratio, the secondary can be modelled through a `skeletonized' source approach \cite{Mathisson1937,PointParticle}:
{\setlength{\abovedisplayskip}{6pt}
 \setlength{\belowdisplayskip}{6pt}
\begin{equation}
    T^{\mu\nu}[\bar{g}]=m_p\int u^\mu_pu^\nu_p \frac{\delta^{(4)}(x^\mu-x^\mu_p(\tau))}{\sqrt{-\bar{g}}}d\tau
\end{equation}
}
with $x^{\mu}_p$ a geodesic in the metric $\bar{g}$.
For a superradiantly formed cloud, the effects can be studied in the linear regime \cite{Validity_expansion}. We can expand the fields in both the mass ratio $q$ and a small parameter $\epsilon \ll 1$ that parameterises the scalar field amplitude, where we track the order of each perturbation through the superscript \cite{BritoMain}:
\begin{align}
    \bm{\mathrm{g}}_{\mu\nu} &= g_{\mu\nu} + q\,h_{\mu\nu}^{(1,0)}+\epsilon\, h_{\mu\nu}^{(0,1)}+\epsilon q\, h_{\mu\nu}^{(1,1)}+\mathcal{O}(q^2,\epsilon^2) \\
    \bm{\phi} &= \epsilon\, \phi^{(0,1)}+\epsilon q\, \phi^{(1,1)}+\mathcal{O}(q^2,\epsilon^2)
    \label{scal field expansion}
\end{align}
Throughout this work, we consider a spherical background $\epsilon\phi^{(0,1)}$ in the quasi-bound ground state. At zeroth order in the mass ratio, this determines the backreaction of the cloud on the metric. As shown in Refs. \cite{EF-BG,EF-BG2}, in Schwarzschild coordinates the leading backreaction will diverge near the horizon as $h_{\mu\nu}^{(0,2)}|_H\sim 1/(r-2M)$, due to the horizon coordinate singularity. This breakdown of the perturbative regime does not occur in ingoing Eddington Finkelstein coordinates \cite{EF-BG,EF-BG2}, where the backreaction comes in as a regular, small correction:

{\footnotesize
\begin{equation}
    ds^2 = -\left( 1\!-\!\frac{2(M\!+\!\epsilon^2 \delta M)}{r}\right)e^{2\epsilon^2 \delta \lambda}dv^2+2e^{\epsilon^2 \delta \lambda}dvdr+r^2 d\Omega^2
    \label{background}
\end{equation}
}

In spherical symmetry, the cloud backreaction is fully described by two functions $\delta M(v,r)$, $\delta\lambda(v,r)$ \cite{EF-BG,EF-BG2}. Following Ref. \cite{BritoMain}, we ignore the slow decay of the cloud $|M\Im(\omega)|\ll 1$, assuming it remains negligible on the fast orbital timescale, and as a proxy for what happens when true bound states are superradiantly formed around rotating black holes \cite{BritoMain}. Recent results in the fully rotating setup confirm its validity \cite{Dyson}. The total mass of the cloud is $M_c = \epsilon^2 \delta M(r\to\infty)$, while we choose $\delta\lambda(r\to\infty)=0$. The new metric Eq. \eqref{background} leads to shifts in geodesic energy and frequency (conservative corrections).

\smallskip

\textit{\textbf{Perturbation Equations.}} At leading dissipative order $\sim \mathcal{O}(q,\epsilon^2)$, the Einstein and Klein Gordon equations are \cite{BritoMain}:
\begin{align}
    &\delta G_{\mu\nu}[\bar{g},h]= 8\pi\, T_{\mu\nu}[\bar{g}]\nonumber\\
    &+ 8\pi \epsilon^2 \left(T_{\mu\nu}^{\phi(2)}[\phi^{(0,1)},\phi^{(1,1)}{}^*] +T_{\mu\nu}^{\phi(2)}[\phi^{(0,1)}{}^*,\phi^{(1,1)}]\right) \nonumber\\
    &+8\pi\epsilon^2 S_{\mu\nu}^h[h,\phi^{(0,1)},\phi^{(0,1)}{}^*] \label{EinsEq} \\[7pt]
    &(\Box^{g}-\mu^2)\phi^{(1,1)} = \,S^{\phi}[h,\phi^{(0,1)}]
    \label{scalEq}
\end{align}

with the line element for $\bar{g}$ given by Eq. \eqref{background}. The $\mathcal{O}(q^1)$ metric perturbation $h$ will contain both $\mathcal{O}(\epsilon^0)$ vacuum and $\mathcal{O}(\epsilon^2)$ environmental terms. Importantly, for the dynamics of the EMRI system, both the scalar field perturbations and $\mathcal{O}(\epsilon^2)$ metric perturbations come in at the same order. This metric backreaction has previously been studied in the axial sector, where the metric perturbations $h^{(1,2)}$ decouple from the scalar $\phi^{(1,1)}$, and a single master equation can be derived \cite{BritoMain}. In this work, we treat the fully coupled and dominant polar sector. We solve Eqs. \eqref{EinsEq}, \eqref{scalEq} in Zerilli gauge, with the scalar profile in a spherical background. We can expand Eq. \eqref{scal field expansion} as:
\begin{equation}
    \phi = \frac{e^{-i\omega v}}{\sqrt{4\pi}}\phi_0(r) + q \sum_{l,m}\frac{e^{-i\omega v}}{r}Y_{lm} \cdot\phi^+_{lm}(v,r) +\mathcal{O}(q^2\epsilon)
\end{equation}
For the conjugate field $\phi^*$, the complex conjugate of the spherical harmonic $Y_{lm}^*$ will appear. The Einstein equations Eq. \eqref{EinsEq} can be decomposed into spherical harmonics again by introducing the function $\phi^-_{lm}(v,r)=(-1)^{m}\phi^{+,*}_{l,-m}(v,r)$ \cite{OG_scalarPert,BritoMain, Duque_2024}, showing a full coupling between $m\!>\!0$ and $m\!<\!0$ scalar perturbations with the metric. Extending the framework developed by Zerilli to compute polar metric perturbations on a Schwarzschild metric
\cite{Zerilli,OG_scalarPert}, Eq. \eqref{EinsEq} can be simplified into two first order equations for generic metrics $ds^2=-A(r)dv^2+C(r)dvdr+r^2d\Omega^2$. We choose to work with the functions $K(r)$ and $H_K(r) = (K(r)+H_1(r))/r$, as this combination of the Zerilli variables somewhat simplifies the equations. Only for a vacuum Schwarzschild black hole, the two equations can be further reduced to a single, second order equation for the Zerilli variable $\psi_{\text{pol}}$. With the asymptotically flat metric Eq. \eqref{background} and exponentially decaying field $\phi_0$, the equations still decouple at infinity. This allows a computation of the infinity flux through $\psi_{\text{pol}}(K,H_K)$ \cite{BritoMain, Cardoso_2022, Infinity_flux}, with:
\begin{equation}
    \dot{E}_\infty^{\text{pol/ax}} = \lim_{r\to\infty}\frac{1}{64\pi}\sum_{l,m}\frac{(l+2)!}{(l-2)!}\left|\dot{\psi}^{lm}_{\text{pol/ax}}\right|^2
    \label{fluxEq}
\end{equation}
Away from resonances, corrections to the horizon flux remain subdominant, and do not affect the main conclusions from this work. In the modes $l\!\geq\!2$, scalar field resonances only appear at radii $r_p\!\geq\! 50M$, and for circular orbits $\dot{E}_H^{\text{GW}}/\dot{E}_{\infty}^{\text{GW}} = \mathcal{O}(r_p^{-4})$ \cite{BritoMain,Horizon_flux}.

\smallskip

\textit{\textbf{Numerical Procedure.}} We solve for the scalar background in terms of Confluent Heun functions \cite{Fiziev_2006}, adapted to include a scalar mass $\mu$. With these solutions, the roots of the quasi bound frequency can be solved in a similar way to Ref. \cite{Fiziev_2006}. In the metric Eq. \eqref{background}, corrections to the geodesic parameters are given in Ref. \cite{BritoMain, Cardoso_2022}. Axial perturbations can be solved through a single second order master equation. We use the variation of parameters approach (see Ref. \cite{BritoMain}). %, correcting for a tiny $1/C(r)$ term. 
For the homogeneous solution at infinity, we derive the solution in two parts. One part solves for a constant mass scaling $M\to \gamma M$, with $\gamma=1+M_c/M$, through a simple rescaling of the vacuum solution $(r\to r/\gamma,\sigma \to \gamma \sigma)$. The remainder can be solved by taking out the exponential damping factor in each of the terms, and explicitly integrating the analytic part. We found this split up to give better numerical accuracy, especially for the more involved polar sector. 

\medskip

In the polar sector, the dynamical fields $(K,H_K,\phi^+,\phi^-)$ are determined as a fully coupled system through Eqs. \eqref{EinsEq} and \eqref{scalEq}. We follow the methods outlined in Refs. \cite{Integration_Method,OG_scalarPert}. The polar equations can be written as:
\begin{equation}
    \frac{d\bm{\Psi}}{dr}+\bm{V}\bm{\Psi}=\bm{S}
    \label{polar eq method}
\end{equation}
where we introduced the six-dimensional vector $\bm{\Psi}=(K,H_K,\phi^+,\phi^+{}',\phi^-,\phi^-{}')$. Matrix $\bm{V}$ is a function of the background fields $(\phi_0,\phi_0^*,\delta M,\delta\lambda)$. In this formulation, the source vector $\bm{S}$ only has support at the position of the secondary $r_p$. We build the particular solution through six independent homogeneous solutions: three ingoing at the horizon, three outgoing at infinity  \cite{Integration_Method,OG_scalarPert}. With the point-particle source, the source integral can be performed analytically. We compute four independent solutions with $(K,H_K)=\mathcal{O}(\epsilon^2)$, where each scalar perturbation $(\phi^+_{H,\infty}$, $\phi^-_{H,\infty})$ can be solved as a vacuum equation (without $\mathcal{O}(\epsilon)$ source term). Solutions $\phi^-_{lm}$ are computed from $\phi^+_{l,-m}$. For the two other solutions, $(K,H_K)=\mathcal{O}(\epsilon^0)$, both scalar perturbations \hbox{$(\phi^+$, $\phi^-)$} are sourced by the metric perturbations, to then become an additional source for the $(K,H_K)=\mathcal{O}(\epsilon^2)$ terms in Eqs. \eqref{EinsEq}. At infinity, we can take out the constant mass shift, and the exponentially damped prefactors as discussed before. For better numerical stability, we solve the vacuum metric perturbations by computing $\psi_{\text{pol}}$, and then use $K(\psi_{\text{pol}},\psi_{\text{pol}}')$, $H_K(\psi_{\text{pol}},\psi_{\text{pol}}')$. Integrations are performed with built-in \texttt{Mathematica} finite difference solvers. For $\phi^-_\infty$, we use a pseudospectral method, on Chebyshev Gauss–Lobatto collocation points, with $N=600$. This was found to be essential in finding the regular sourced solution, with an exponentially growing homogeneous solution leaving the finite difference method unstable (see also Appendix \ref{App E}). By evaluating each of the six solutions back into the field equations Eq. \eqref{EinsEq}, we could verify that the error indeed scales $\sim M_c^2$. Finally, since we expand the field equations up to first order in $M_c$, any terms $\mathcal{O}(M_c^2)$ in the flux are unphysical. To remove these, we compute the flux at three points $M_c=(0.001,0.002,0.003)M$, and extrapolate out the higher order terms (see Appendix \ref{App B}). Vacuum fluxes are computed with the \texttt{ReggeWheeler} package of the \texttt{Black Hole Perturbation Toolkit} \cite{BHPToolkit}, and agree with our fluxes for $M_c=0$ to $\sim 10^{-10}$, indicating very good agreement. In this approach, scalar and polar fluxes are jointly computed within the coupled system, finding the scalar fluxes to be consistent with Ref. \cite{BritoMain}. If not shown explicitly, scalar perturbations are summed up for all modes $l=1,2,3$, $m=\pm l$, (consistent with Ref. \cite{BritoMain}), polar perturbations for $l=2,3$, $m=\pm l$, axial perturbations for $l=2,3$ and $m=\pm1,\pm2$ respectively.

\begin{figure}[ht]
    \centering
    \includegraphics[width=0.97\linewidth]{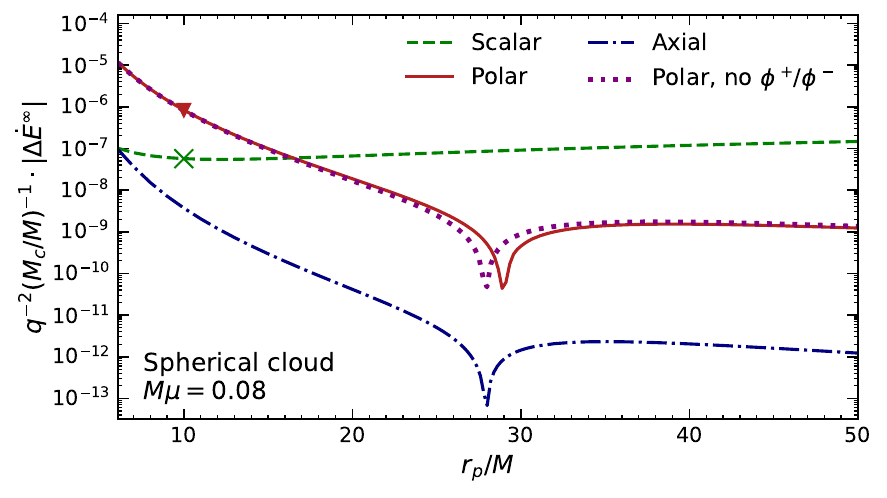}
    \setlength\abovecaptionskip{-4pt}
    \setlength\belowcaptionskip{-6pt}
    \caption{Cloud corrections to the vacuum flux for each of the terms in Eq. \eqref{flux term Eq}, with $M\mu=0.08$. At small (large) radii, axial and polar terms give a negative (positive) correction. The `Polar, no $\phi^+/\phi^-$' curve is computed with $\phi^+=\phi^-=0$. The fluxes at $r_p=10M$ are indicated as a triangle and cross, and are shown in Fig. \ref{fig: mass scaling} as well.}
    \label{fig: flux terms}
\end{figure}

\textit{\textbf{Flux contributions.}}
The total, radiative flux at infinity is given by \cite{BritoMain,Dyson}:
\begin{equation}
    \dot{E}^\infty = q^2\dot{E}_{\text{GW}}^{\text{vac}} + q^2\epsilon^2(\Delta\dot{E}_{\text{polar}}+\Delta\dot{E}_{\text{axial}}+\dot{E}_{\text{scal}})
    \label{flux term Eq}
\end{equation}
For the first time, we present the relative contribution of each of these terms in a scalar environment (Fig. \ref{fig: flux terms}, with $M\mu = 0.08$). Remarkably, the polar flux is non-negligible, and even highly dominant at low radii\footnote{Interestingly, similar conclusions were found in Ref. \cite{POSTADIABA}, in another environment with a dense, Hernquist/ NFW type dark matter halo, and vanishing radial pressure. }, right in the frequency range for measurements with LISA. The main features are:

(i) At small separation, the polar correction to the flux is negative. In the limit of low compactness, its value matches increasingly well with a redshift (see further), similar to the axial sector (see Appendix \ref{App C}, and Ref. \cite{BritoMain}). (ii) At large separation, the polar correction increases the flux. In the limit of $r\!\to\! \infty$, the secondary only feels the mass potential of the cloud, and the flux is determined by a simple mass shift $M\to M+M_c$. For high compactness $M\mu=0.3$, we can indeed recover the expected scaling (see Appendix \ref{App D}). (iii) At an intermediate radius, the flux transitions from negative to positive, with the precise turnover determined through the full integration of the field equations Eq. \eqref{EinsEq}. With increasing mass $M\mu$, the location of the turnover point switches to the left, closely resembling the axial sector, and almost reaching the ISCO for $M\mu=0.2$. For these heavier $M\mu$, the scalar flux remains dominant. (iv) To study the coupling to scalar perturbations, we temporarily set the scalar perturbations to zero ($\phi^+=\phi^-=0$), greatly simplifying the problem by requiring only two homogeneous solutions for Eq. \eqref{polar eq method}. Scalar perturbations only weakly couple, and can almost be ignored for most of the parameter space. The strongest deviation arises at the turnover point, demonstrating that its location is non-trivially set through the integration.

\begin{figure}
    \centering
    \includegraphics[width=0.97\linewidth]{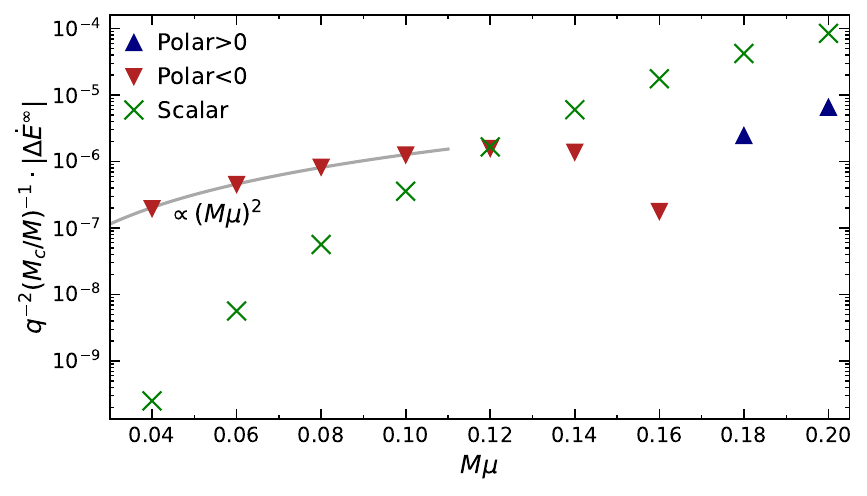}
    \setlength\abovecaptionskip{-4pt}
    \setlength\belowcaptionskip{-10pt}
    \caption{Scalar and polar flux corrections as a function of $M\mu$, at a fixed radius $r_p=10M$. Down-pointing triangles in red show a negative flux correction (redshift), up-pointing arrows in blue show a positive correction (blueshift/mass shift). Scalar fluxes are always positive for a spherical cloud \cite{BritoMain}. At small mass $M\mu$, we find a good quadratic fit for the polar term.}
    \label{fig: mass scaling}
\end{figure}

\smallskip

\textit{\textbf{Scalar mass scaling.}}
To examine the scalar mass range where $\Delta \dot{E}_{\text{polar}}$ becomes dominant, we extract the polar and scalar flux terms at a fixed radius $r_p\!=\!10M$, as a function of $M\mu$ (Fig. \ref{fig: mass scaling}). In the LISA band, a realistic EMRI system can spend its entire observation time at separations $r_p \lesssim 10M$ \cite{Khalvati_2025,FEW1}. At these separations relevant for LISA, we find that the polar flux is dominant at low masses $M\mu \!\lesssim\! 0.12$. At smaller radii $r_p\!<\!10M$, scalar fluxes only grow slightly (Fig. \ref{fig: flux terms} and Ref. \cite{BritoMain}), while the polar correction approximately grows with the vacuum flux, further enhancing the scaling between the two. Note that at the mass $M\mu=0.12$, both contributions almost exactly cancel out. Between $M\mu=0.16$ and $M\mu=0.18$, the positive/negative turnover point is at $r_p=10M$. For heavier masses $M\mu\gtrsim0.18$, the polar correction increases the vacuum flux, and remains non-negligible.

\smallskip

\begin{figure}
    \centering
    \includegraphics[width=0.955\linewidth]{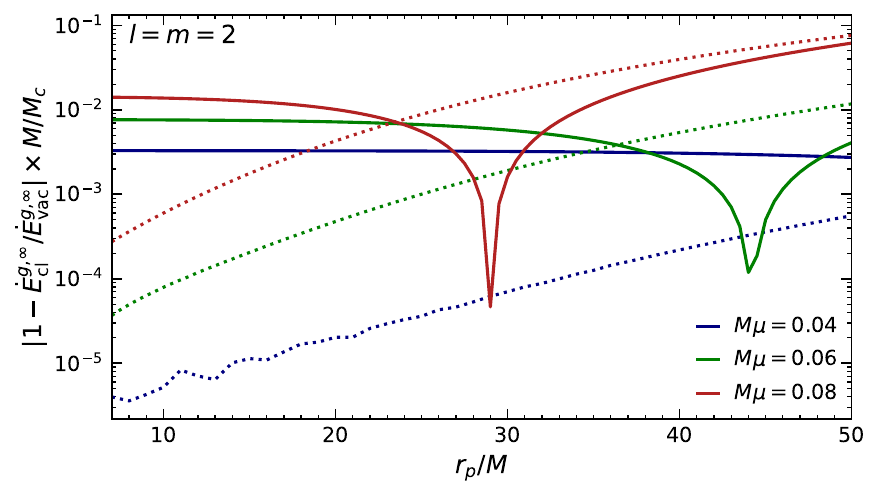}
    \setlength\abovecaptionskip{-0.6pt}
    \setlength\belowcaptionskip{-6pt}
    \caption{Relative difference between the $l\!=\!m\!=\!2$ vacuum infinity flux and the flux with cloud coupling (solid line), for clouds with low compactness. Dotted lines correspond to the relative difference with the redshifted flux.}
    \label{fig: redshift}
\end{figure}

\textit{\textbf{Flux redshift.}}
For the regime where the polar term can become dominant, we assess the detectability of this new correction in the waveform. We can take inspiration from the axial sector, where a simple master equation can be derived. At small radii, the axial master equation expands into a simple redshift for the flux (see Appendix \ref{App C}, and Ref. \cite{BritoMain}):
\begin{equation}
    \Delta\dot{E}_{\text{axial}}^{lm}(r_p) = 2\delta \lambda_H\left(\frac{M_c}{M}\right) \dot{E}_{\text{axial}}^{lm,\text{vac}}(r_p)
    \label{redshift Eq}
\end{equation}
with $\delta\lambda_H = \delta\lambda(r\to2M)$. In the limit of small $M\mu$, $\delta\lambda_H\sim (M\mu)^2$ \cite{BritoMain}. For a different EMRI system in a dark matter halo with vanishing radial pressure, Ref. \cite{Cardoso_2022} shows that the coupling to the fluid breaks this redshift degeneracy in the polar sector. However, in a scalar background the metric only weakly couples to the scalar perturbations (see Fig. \ref{fig: flux terms}). Additionally, at low $M\mu$ and low radius $r_p$, the polar flux approximately scales as $\propto\! (M\mu)^2$ (see Fig. \ref{fig: mass scaling}), as expected for a redshift. In Fig. \ref{fig: redshift}, we present the relative difference of the cloud flux with vacuum, and with fluxes redshifted according to Eq. \eqref{redshift Eq}. In a consistent result, we conclude that even in the polar sector, the fluxes match increasingly well with a simple redshift as $M\mu \ll 1$. 

%Given a correspondence between minimally coupled scalar fields and effective perfect fluids \cite{Faraoni_2012}, the different conclusions from Ref. \cite{Cardoso_2022} may at first appear surprising. However, they can be explained by our different setup: the source contains radial pressure, the infinity profile follows a mass shift $M\!\to\!M\!+\!M_c$ and not a Hernquist profile, the superradiant cloud is in a different regime where $M_c\ll M$, and we assume different boundary conditions for the perturbations, physically motivated through the Klein Gordon equation (ingoing/outgoing vs. Dirichlet conditions).

\begin{figure}
    \centering
    \includegraphics[width=1\linewidth]{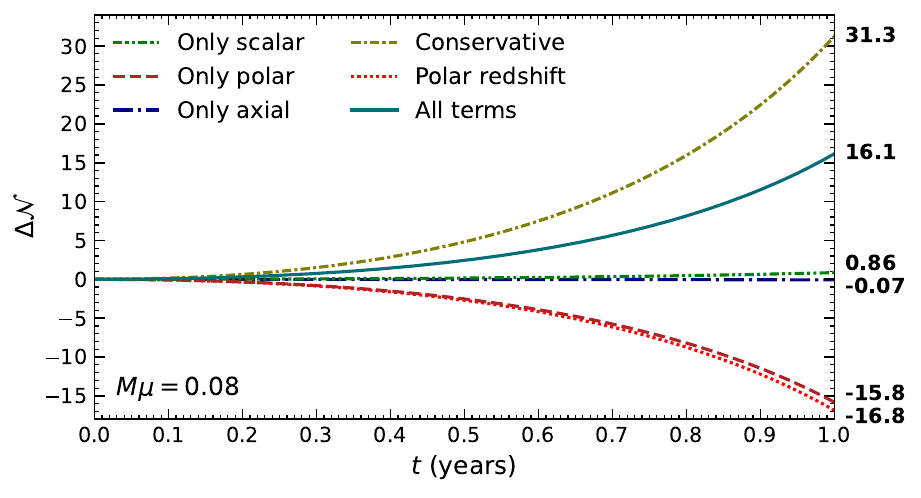}
    \setlength\abovecaptionskip{-11.7pt}
    \setlength\belowcaptionskip{-20pt}
    \caption{Number of cycles dephasing $\Delta \mathcal{N}$ with the vacuum waveform, for $M=M_1=10^6M_\odot$, $M_2=10^1M_\odot$, $r_0 = 10M$, $M\mu=0.08$, $M_c=0.1M$. When including conservative corrections, we slightly reduce the radius $r_0 \approx 9.995M$ to make the initial frequencies match. The total number of cycles is $\sim 4\cdot10^5$ after one year.}.
    \label{fig: dephasing}
\end{figure}

\smallskip

\textit{\textbf{Dephasing.}}
To assess the impact on the waveform, we compare the dephasing induced by each cloud correction in the relativistic regime. The method is outlined in Appendix \ref{App F}. Vacuum GW fluxes are supplied by the \texttt{Black Hole Perturbation Toolkit} \cite{BHPToolkit}. With $M_c=0$, the dephasing with respect to the \texttt{FEW} \cite{FEW2} trajectory is $\mathcal{O}(10^{-6})$ cycles, indicating good agreement. The orbital phase is multiplied by a factor 2, to match the dominant GW phase in the $m\!=\!2$ modes. Conservative corrections arise from shifts in the geodesic energy $\delta E$ and frequency $\delta\Omega$, through the backreaction of the cloud on the metric \cite{BritoMain,Cardoso_2022}. We consider a representative EMRI system\footnote{We use the new \texttt{FEW} mass conventions, see Appendix A of \cite{FEW2}.}, with $M=M_1=10^6M_\odot$, $M_2 = 10^1 M_\odot$, $r_0=10M$, $M\mu=0.08$. \cite{FEW1,FEW2,Khalvati_2025}. For the mass of the cloud, we consider the idealistic case $M_c=0.1M$ \cite{Superradiance, BritoMain}. In more realistic scenarios, the evolution of the orbit through resonances will lower this upper limit, scaling the dephasing \cite{scal13,scal14}. The phase with conservative corrections is started at a slightly smaller radius to match the vacuum frequency. We assess the importance of including each of these cloud signatures in the waveform, shown in Fig. \ref{fig: dephasing}. We note several key findings:

(i) Conservative terms make up the dominant cloud correction, for this set of parameters. In the limit $M\mu \ll 1$, $\delta M\!\to\!0$ and $\delta \lambda \!\to \!\delta \lambda_H$ at small separations. Due to the expressions for $\delta E$ and $\delta\Omega$, we can not match both energy and frequency with its vacuum values, resulting in a dephasing. Additionally, the shifts $\delta E$ and $\delta \Omega$ change with separation. For small $M\mu$, this scales with the magnitude of $\delta \lambda_H \sim (M\mu)^2$, further dominating the scalar flux corrections at smaller $M\mu$.
(ii) The polar correction, from this work, gives the dominant dephasing arising from the fluxes. We can approximate the dephasing by a simple shift in the mass ratio through Eq. \eqref{redshift Eq} (flux redshift).
(iii) Axial corrections are almost negligible. (iv) For this set of parameters, scalar radiation is also highly subdominant. (v) After 1 year, the dephasing for this light $M\mu$ is $\approx 16$ cycles. Closer to the ISCO, the dephasing will rapidly increase, and polar/conservative corrections will further dominate the scalar radiation channel (see \hbox{Fig. \ref{fig: flux terms}}). (vi) We further analyse the dephasing for heavier $M\mu=0.2$ (see Appendix \ref{App G}). The polar correction is subdominant, but still non-negligible. Interestingly, conservative corrections can become dominant even in this higher $M\mu$ regime. Instead of a redshift, at large radii the vacuum geodesic energy $E = m_p(1-M/(2r)+\mathcal{O}(1/r^2))$ is mainly adjusted by an increased mass $M\!\to\! M\!+\!M_c$, deepening the potential and increasing the binding energy. With more energy to radiate, the conservative shift slows down the inspiral. Even at a lower radius $r_0=10M$, it can almost exactly cancel the increased flux from the scalar channel (see Appendix \ref{App G}). However, when degenerate with a mass shift, it should not significantly alter detectability of the cloud. Only in the intermediate regime where conservative effects are neither a mass nor redshift, one could expect a significant change in cloud parameter estimations. A comprehensive parameter estimation analysis will be needed to fully quantify this.

\medskip

\textit{\textbf{Discussion.}} 
In this work, we have shown that the fully coupled polar sector can give the dominant dissipative cloud correction, a signature that was previously missing in the literature. We provide a characterisation of the flux profile from its two limits: an approximate redshift at close, and a mass shift at large separation. In the spherical cloud parameter space, we identify $M\mu \lesssim 0.12$ as the regime where polar overtake axial and scalar flux channels, while remaining non-negligible in the higher $M\mu$ regime. %As a new cloud correction, it is redshift dominated at low $M\mu$, mostly degenerate with the mass ratio. 
%Whether the combination with conservative effects, giving a dephasing in the other direction, could lift this parameter degeneracy remains to be studied. 
On the level of the waveform, we showed that conservative corrections can further dominate the total dephasing. Interestingly, for more compact clouds studied in \cite{BritoMain,Duque_2024,Dyson,scalarRadiation,Khalvati_2025}, conservative effects can nearly cancel the scalar flux. A comprehensive parameter estimation analysis will be needed to fully quantify how this could affect the inference. While these methods are applied to a simplified setup (spherical cloud, non-spinning primary, circular orbits), we expect the general conclusions to still hold. Monopole corrections from a dipole cloud have been shown to give a similar metric redshift function \cite{Redshift_dipole}. With Newtonian expressions for the dipole profile (Ref. \cite{Newtonian_expr,Detweiler}), this redshift still scales quadratically with the field mass, indicating that it could again dominate scalar radiation at small $M\mu$. If dipole and quadrupole contributions break the simple redshift degeneracy, this could excitingly increase the dipole detectability towards lower scalar masses. In the high $M\mu$ regime, the polar correction should still be understood as a mass shift at large radii. For a spinning primary, Refs. \cite{Dyson,scalarRadiation} show that spin effects mostly come in at more relativistic, higher $M\mu$. This would suggest that polar corrections can still dominate in the lower $M\mu$ regime. Our results motivate extending the analysis of polar and conservative cloud corrections to more generic configurations, and indicate the importance of incorporating these effects in beyond‑vacuum EMRI templates. Finally, there are still other important issues that remain to be answered. An evolution of the environmental system through only radiative fluxes has recently been questioned in Refs. \cite{ConorTorque,Macedo}, where non-radiative torques will come in as an additional dephasing. Corrections to the cloud eigenfrequency additionally contribute to the non-radiative modes, scalar horizon fluxes get corrections from terms $\phi^+\phi_0\psi_{\text{pol}}$ (see Ref. \cite{BritoMain}). We plan to address these questions in future work.

\medskip

\textit{\textbf{Acknowledgements.}} 
We are indebted to Richard Brito, Vitor Cardoso, Conor Dyson and Thomas Spieksma for their insightful comments on the manuscript. We would further like to thank Bert Depoorter, Llibert Aresté Saló and Tom van der Steen for valuable discussions and suggestions, and Archisman Ghosh for the guidance of R.K. during his master thesis on the topic. The authors thank the Belgian Federal Science Policy Office (BELSPO) for the provision of financial support in the framework of the PRODEX Programme of the European Space Agency (ESA) under contract number PEA4000144253. This research was supported in part by KU Leuven grant C16/25/010.
% S.M.
We acknowledge support by VILLUM Foundation (grant number VIL37766). The Center of Gravity is a Center of Excellence funded by the Danish National Research Foundation under grant number DNRF184. This research was supported by Simons Foundation International and the Simons Foundation through Simons Foundation grant SFI-MPS-BH-00012593-11.

\bibliographystyle{apsrev4-2}
\bibliography{RelativisticSignatures}

\newpage

\section*{Appendices}

\subsection{Units}
\label{App A}
The conversion between Planck units and physical units is given by \cite{Superradiance}:
\begin{align}
    \mu = \alpha \left(\frac{\hbar c}{G}\right)\frac{1}{M}
    \approx \alpha\,(1.34\cdot 10^{-10} \text{ eV/c$^2$})\left(\frac{M_\odot}{M}\right)
\end{align}

Using EMRIs and IMRIs, previous studies have estimated the masses detectable in the LISA frequency band to be in the range $\mu \in [10^{-16.5},10^{-14}]$ eV, or generally $\alpha=M\mu \in[0.05,0.3]$ \cite{Tjonnie,BritoMain,Khalvati_2025}.

\subsection{Boundary conditions}
\label{App E}

At infinity, the homogenous solutions for the dynamical fields will be approximately of the form:
\begin{align}
    K(r)_{r\to \infty} &= (\zeta_{k_0}^{\infty}(r)+\epsilon^2 \zeta_{k_2}^{\infty}(r))\sum_{i=0}^{N}\frac{(k_i^{(0)}+\epsilon^2 k_i^{(2)})}{r^i} \\
    H_K(r)_{r\to \infty} &= (\zeta_{h_0}^{\infty}(r)+\epsilon^2 \zeta_{h_2}^{\infty}(r))\sum_{i=0}^{N}\frac{(h_i^{(0)}+\epsilon^2 h_i^{(2)})}{r^i} \\
    \phi^+_{r\to \infty} &= \zeta_{\phi^+}^{\infty}(r)\sum_{i=0}^{N}\frac{p_i}{r^i} \\
    \phi^-_{r\to \infty} &= \zeta_{\phi^-}^{\infty}(r)\sum_{i=0}^{N}\frac{m_i}{r^i}
\end{align}
where we split up the solution into non-analytic factors $\zeta^{\infty}(r)$ and analytic factors $\sum a_i/r^i$. Depending on the field, we include at least three terms up to $N=2$. At the horizon, the solutions $\zeta^{H}(r)\cdot\sum a_i(r-2M)^i$ all have trivial `non-analytic' factors:
\begin{equation}
    1=\zeta_k^H=\zeta_h^H=\zeta_{\phi^+}^H=\zeta_{\phi^-}^H
\end{equation}
when working in ingoing Eddington Finkelstein coordinates. The factors $\zeta^{\infty}(r)$ can be found by iterating through the expansion: the lowest order field is always outgoing/diminishing at infinity, which then acts as a source for higher order fields, when multiplied with a background field $(\phi_0,\phi_0^*,\delta M,\delta\lambda)$. Except for a constant mass shift $\delta M(r\to\infty) = M_c$, which can be solved by a rescaling of the vacuum coordinates, these background fields exponentially decay to zero. We take out the non-analytic factors $\zeta^\infty(r)$, and integrate the remainder. The correct boundary conditions are imposed through the series expansions, evaluated at $10^{10}M$, or $10^7M$ for the higher coupled fields at infinity, or $(2+10^{-8})M$ at the horizon. We verified that the results are unchanged when varying this infinity and horizon limit. In the pseudospectral solution for $\phi^-_\infty$ with source, we compactify the range $[6M,10^7M]$ to $[-1,1]$, by a reparametrisation of the form: $-1 = a+b/(10^7M)$, $1=a+b/(6M)$. With the pseudospectral method, we verified that the value at $10^7M$ matches the series solution to within $\sim10^{-4}$, confirming that the regular solution found by the pseudospectral method is indeed the series solution above.

\subsection{Extrapolation of higher order terms.}
\label{App B}

\begin{figure}[H]
    \centering
    \includegraphics[width=1\linewidth]{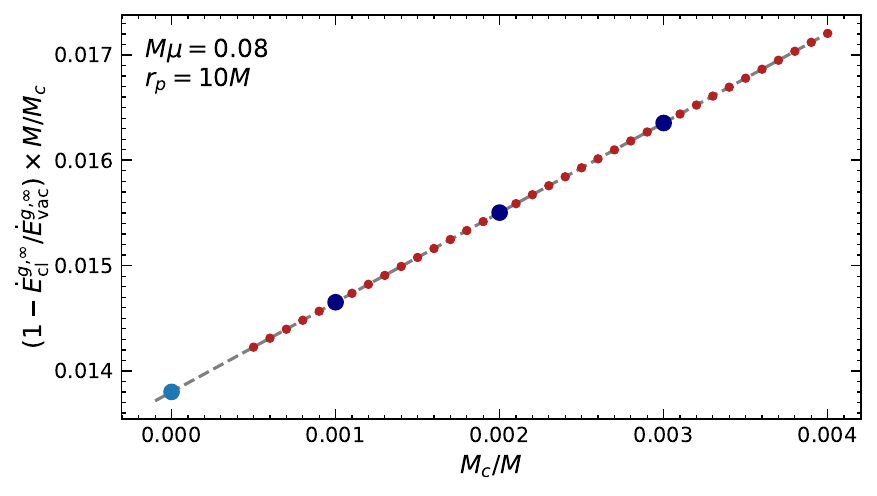}
    \caption{Extrapolation of higher order terms. Unphysical $\mathcal{O}(M_c/M)$ terms are significant, but we have good control over the error (linear). Red dots show the $l=m=2$ flux recovered for each mass $M_c/M$, dark blue dots show the standard evaluation points $M_c=(0.001,0.002,0.003)M$. Through extrapolation, we can accurately recover the $\Delta\dot{E}/\dot{E}_{\text{vac}}$ correction (light blue dot).}
    \label{fig: cloud mass}
\end{figure}

The polar energy flux at small mass $M_c$ can be expanded as:
\begin{equation}
    \dot{E} = \dot{E}_{\text{vac}} + (M_c/M)\Delta\dot{E} + \mathcal{O}((M_c/M)^2)
\end{equation}
where any $\mathcal{O}((M_c/M)^2)$ terms in our result are by definition non-physical, due to the expansion of the field equations Eq. \eqref{EinsEq} and Eq. \eqref{scalEq} \textit{a priori}. We can assess the magnitude of the expansion error by computing $(1-\dot{E}/\dot{E}_{\text{vac}})(M/M_c) = -\Delta\dot{E}/\dot{E}_{\text{vac}} + \mathcal{O}(M_c/M)$. For the masses where we compute the flux $M_c=(0.001,0.002,0.003)M$, it is clear that they can not be neglected, when demanding at least a two digit accuracy on the flux correction (see Fig. \ref{fig: cloud mass}). However, we have a good control over the expansion error at these small masses: only the first term $\mathcal{O}(M_c/M)$ is significant, with a good linear fit. This allows an extrapolation of the flux correction $\Delta\dot{E}$ by the value of the linear fit at $M_c=0$. To assess the extrapolation error, we always compute a third point at $M_c=0.003M$, allowing a quadratic fit. The difference in $\Delta\dot{E}/\dot{E}_{\text{vac}}$ for the linear and quadratic fit is well below $10^{-2}$ for all fluxes in this work, achieving the desired precision, and confirming the stability of the numerical integrations.

\subsection{Axial flux, close separation and low compactness}
\label{App C}

\begin{figure}
    \centering
    \includegraphics[width=1\linewidth]{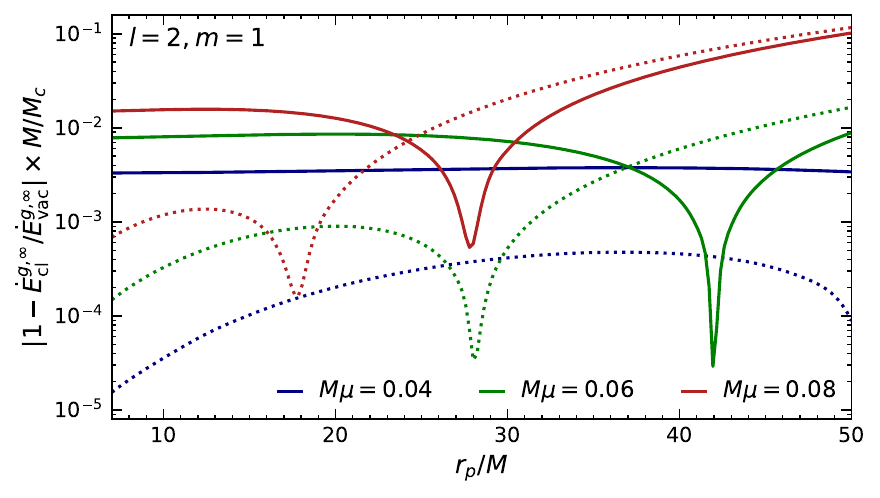}
    \setlength\abovecaptionskip{-10pt}
    \caption{Limit of close separation, low compactness, axial sector. Solid (dotted) lines correspond to the relative difference with the vacuum (redshifted) flux.}
    \label{fig: axial redshift}
\end{figure}

At close separation and low compactness, the background functions quickly converge as $\delta M\to 0$, $\delta \lambda \to \delta \lambda_H$, and the metric simplifies to a Schwarzschild black hole with a redshift on the time variables $v$. In this section, we would like to clear up some subtleties in the literature, and derive the redshift used for the axial sector. With the master variable from Ref. \cite{BritoMain}, the axial master equation in this limit becomes:
\begin{equation}
    \left[ \frac{d^2}{d(r_*^{\text{vac}})^2}+\frac{\sigma^2}{\gamma^2}-V_{\text{ax}}^{\text{vac}}\right]\bar{\psi}_{\text{ax}}^{lm}(r) = S_{\text{ax}}^{\text{vac},lm}(r)
    \label{axial master eq}
\end{equation}
where $\gamma = (1+\epsilon^2\delta\lambda_H)$. We correct a missing factor $\gamma$ in Ref. \cite{BritoMain} for the source function, which does correctly appear for the different master variable used in Ref. \cite{Cardoso_2022}. Only at the level of the fluxes, both definitions match, through the multiplication by another $\sigma$. We can further simplify Eq. \eqref{axial master eq} in terms of the \textit{vacuum} circular orbit frequency $\sigma_0 = m \sqrt{M/r_p^{3}}=(\sigma/\gamma)|_H$, where the last equality holds at the horizon. With $\sigma_0=\sigma/\gamma$, the master equation looks identical to the vacuum case. At the level of the fluxes (see Eq. \eqref{fluxEq}), one can see that redshifted fluxes are simply multiplied by $\gamma^2<1$ (see Fig. \ref{fig: axial redshift}). 

\vspace{20pt}

\subsection{Large separation and high compactness}
\label{App D}

At large separation and high compactness, the background functions quickly converge as $\delta M\to M_c$, $\delta \lambda \to 0$, $\phi_0\to 0$, and the spacetime simply becomes a mass shifted Schwarzschild black hole $M\to M+M_c$. We define $\dot{E}_q=\dot{E}/q^2$ as the flux normalized by the mass ratio. For a small mass shift $M+M_c$ the total flux is:
\begin{equation}
    \dot{E} = q^2 \dot{E}_q = \left( \frac{m_p}{M+M_c}\right)^2 \dot{E}_q = \left( \frac{m_p}{M}\right)^2 \left(1-2\frac{M_c}{M}\right)\dot{E}_q
\end{equation}
The flux is lowered by $(M/m_p)^2\dot{E} = (1-2M_c/M)\dot{E}_q$, due to a diminished mass ratio. However, the radius also remains fixed as $r_p/M$ (not $r_p/(M+M_c)$), introducing a different scaling with mass for each harmonic. The leading, Post-Newtonian flux radiated by GWs, at large separations $r_p$, scales as: $\dot{E}^{\text{polar}}_{r\gg6M} \sim r_p^{-3-l}$ and $\dot{E}^{\text{axial}}_{r\gg6M} \sim r_p^{-4-l}$ \cite{EricPoisson}. When keeping the geodesic at a fixed radius $r_p/M$, the flux for a mass shifted black hole, ignoring the mass ratio, then scales as:
\vspace{-4pt}    
\begin{align}
    \dot{E}^{\text{polar}}_{r\gg6M} &\sim \left(1+(3+l)\frac{M_c}{M}\right) \left(\frac{r_p}{M+M_c}\right)^{-3-l} \\
    \dot{E}^{\text{axial}}_{r\gg6M} &\sim \left(1+(4+l)\frac{M_c}{M}\right) \left(\frac{r_p}{M+M_c}\right)^{-4-l}
\end{align}

Considering both a smaller mass ratio and shift in radius gives the correct scaling with $M_c$ at large radii:
\vspace{-7pt}    
\begin{align}
    \dot{E}^{lm}_\text{polar} &\sim \left[1+(1+l)\frac{M_c}{M}\right]\dot{E}_{\text{vac}}^{lm} 
    \label{lim1}\\
    \dot{E}^{lm}_{\text{axial}} &\sim \left[1+(2+l)\frac{M_c}{M}\right]\dot{E}^{lm}_{\text{vac}}
    \label{lim2}
\end{align}

\begin{figure}
    \centering
    \includegraphics[width=0.96\linewidth]{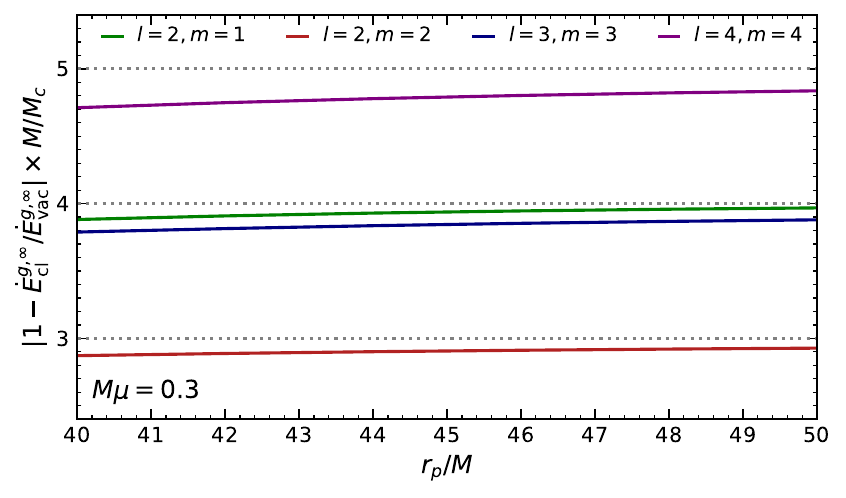}
    \setlength\abovecaptionskip{0.1pt}
    \caption{Limit of large radius, high compactness ($M\mu=0.3$). The expected limits for each of the harmonics are recovered, indicated as grey dotted lines (Eqs. \eqref{lim1}, \eqref{lim2}).}
    \label{fig: inflim}
\end{figure}

We show the scaling factor of four different harmonics in Fig. \ref{fig: inflim}, converging to the expected value. Deviations are expected, both due to $\delta M$ and $\delta \lambda$ not exactly at their limit value, and higher post-Newtonian terms at finite $r_p$.

\subsection{Waveform phase for circular orbits}
\label{App F}

For circular orbits, the orbit is fixed only by a single parameter $r_p$, varying on the slow timescale. We can write: \cite{Dephasing1,Mitra_2025}:

\begin{align}
    \frac{dr_p}{dt} &= \left(q\frac{dE_{\text{orbit}}}{dr}\right)^{-1}\cdot \dot{E}_{\text{orbit}} \\
    \frac{d\phi_p}{dt} &= \Omega(r_p(t))
\end{align}
Assuming we can extend the vacuum energy balance law to an environmental EMRI, we can evolve the trajectory by \cite{BritoMain}:
\begin{equation}
    \dot{E}_{\text{orbit}} = -q^2\dot{E}_{\text{GW}}^{\text{vac}} - q^2(M_c/M)\Delta\dot{E}
\end{equation}
where $\Delta\dot{E}$ contains contributions from the scalar, polar, and axial sector. Note that the energy balance law might have to be updated to include non-radiative torque corrections (as suggested in recent work Refs. \cite{ConorTorque,Macedo}). Any potentially missing terms would come in as additional corrections to our results. Horizon flux corrections are not computed in this work, but remain subdominant \cite{BritoMain,Horizon_flux}, away from resonances with the background. Conservative corrections result from a shift in geodesic energy and frequency $E_{\text{orbit}}\to E^{\text{vac}}(r_p)+(M_c/M)\delta E(r_p)$, $\Omega_{\text{orbit}}\to \Omega^{\text{vac}}(r_p)+(M_c/M)\delta \Omega(r_p)$, computed in \cite{BritoMain, Cardoso_2022}. To make the initial frequencies match with the vacuum waveform, so we can compute a dephasing in the signal, we slightly reduce the initial position to $9.99546537817645M$, for a mass $M_c=0.1M$. With this shift, the initial frequencies match to a relative error of $\mathcal{O}(10^{-17})$. Slightly perturbing this initial separation does not change the conservative dephasing.

\subsection{Dephasing for $M\mu = 0.2$}
\label{App G}

\vspace{-8pt}

\begin{figure}[hb]
    \centering
    \includegraphics[width=1\linewidth]{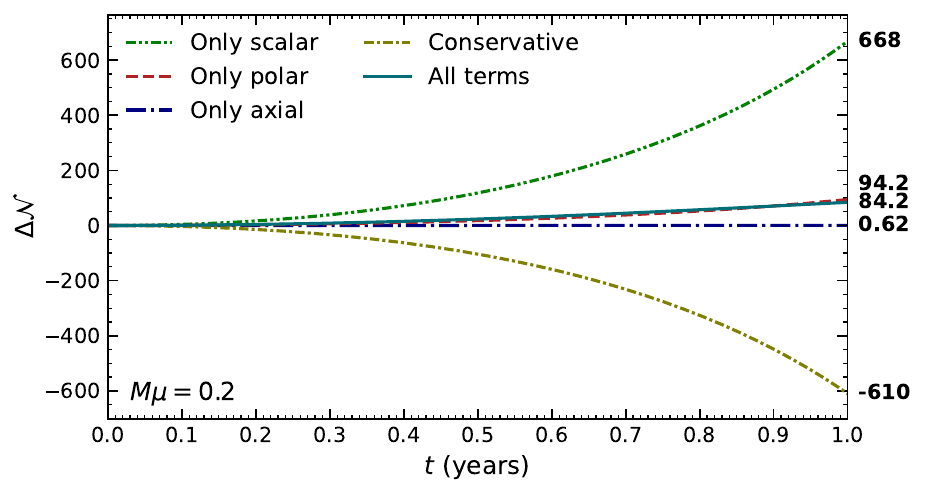}
    \caption{Number of cycles dephasing $\Delta \mathcal{N}$ with the vacuum waveform, for $M=M_1=10^6M_\odot$, $M_2=10^1M_\odot$, $r_0 = 10M$, $M\mu=0.2$, $M_c=0.1M$. When including conservative corrections, we slightly increase the radius $r_0 \approx 10.005M$ to make the initial frequencies match. The total number of cycles is $\sim 4\cdot10^5$ after one year.}
    \label{fig: Mmu02 dephasing}
\end{figure}

For $M\mu=0.2$, the polar correction is always subdominant to the scalar flux. However, conservative effects for these high $M\mu$ will slow down the inspiral at most radii, and almost cancel the increased flux from the scalar channel. Under this cancellation, the polar term can still become highly significant. With conservative effects, we start at a slightly higher radius $10.00466300187813M$, to make the initial frequencies match. Slightly perturbing this initial separation does not change the dephasing. We note that, while this shows the importance of including each term, the exact cancellation of scalar and conservative terms does not imply it should significantly alter parameter estimations. At large radii, conservative corrections mostly represent a simple mass shift $M\!\to\!M\!+\!M_c$, where it would be more informative to compare the dephasing with respect to a mass shifted vacuum solution. Only in the intermediate regime, where it is neither a redshift nor mass shift, one could expect conservative effects to alter cloud parameter estimations. At $r_0=10M$ the dephasing in Fig. \ref{fig: Mmu02 dephasing} presumably \textit{is} in the intermediate regime). Robust conclusions require a thorough parameter estimation analysis.

\end{document}